Chemistry of Atmospheres Formed during Accretion of the Earth and Other Terrestrial Planets

Laura Schaefer

And

Bruce Fegley, Jr.

Planetary Chemistry Laboratory
Department of Earth and Planetary Sciences
Washington University
St. Louis, MO 63130-4899
laura_s@wustl.edu
bfegley@wustl.edu

**Proposed Running Head:** Atmospheric chemistry during planetary accretion

**Corresponding Author:**

Laura Schaefer

Campus Box 1169

Department of Earth and Planetary Science

Washington University

One Brookings Dr.

St. Louis, MO 63130-4899

laura_s@wustl.edu

Phone: 314-935-6310

Fax: 314-935-7361

**Abstract:** We used chemical equilibrium and chemical kinetic calculations to model chemistry of the volatiles released by heating different types of carbonaceous, ordinary and enstatite chondritic material as a function of temperature and pressure. Our results predict the composition of atmospheres formed by outgassing during accretion of the Earth and other terrestrial planets. Outgassing of CI and CM carbonaceous chondritic material produces $H_2O$-rich (steam) atmospheres in agreement with the results of impact experiments. However, outgassing of other types of chondritic material produces atmospheres dominated by other gases. Outgassing of ordinary (H, L, LL) and high iron enstatite (EH) chondritic material yields $H_2$-rich atmospheres with CO and $H_2O$ being the second and third most abundant gases. Outgassing of low iron enstatite (EL) chondritic material gives a CO-rich atmosphere with $H_2$, $CO_2$, and $H_2O$ being the next most abundant gases. Outgassing of CV carbonaceous chondritic material gives a $CO_2$-rich atmosphere with $H_2O$ being the second most abundant gas. Our results predict that the atmospheres formed during accretion of the Earth and Mars were probably $H_2$-rich unless the accreted material was dominantly CI and CM carbonaceous chondritic material. We also predict significant amounts of S, P, Cl, F, Na, and K in accretionary atmospheres at high temperatures (1500 – 2500 K). Finally, our results may be useful for interpreting spectroscopic observations of accreting extrasolar terrestrial planets.

**Keywords:** atmospheric chemistry, planetary accretion, Earth, Mars, extrasolar planets, geochemistry, impacts, atmospheric formation

# 1. Introduction

Earth, Venus, and Mars have secondary atmospheres that originated by the outgassing of volatiles (e.g., CO, $CO_2$, $CH_4$, $H_2$, $H_2O$, $N_2$, and $NH_3$) from minerals during and/or after planetary accretion (e.g., see Brown 1949; Arrhenius et al. 1974; Lange and Ahrens 1982a,b; Abe and Matsui 1985). The outgassed volatiles formed atmospheres on Earth, Mars, and Venus and the Earth's oceans. The atmospheres of extrasolar terrestrial planets may also have formed by outgassing of volatiles during and/or after their accretion.

Earlier we studied outgassing of ordinary chondritic material and its implications for metamorphism on meteorite parent bodies and the composition of Earth's early atmosphere (Schaefer and Fegley 2007). Here we focus on another topic, which is the chemistry of the atmospheres formed on the Earth and other terrestrial planets during their accretion. Our work was stimulated by the concept of a steam atmosphere on the early Earth (e.g., Abe and Matsui 1985). We wanted to use chemical equilibrium calculations to study the chemistry of other gases in the proposed steam atmosphere. Instead we found that atmospheres formed during Earth's accretion were probably $H_2$-rich and that steam atmospheres formed only in special conditions.

Arrhenius et al (1974) calculated that heating during formation of the Earth should release volatiles such as $H_2O$ from the accreted material and lead to formation of an $H_2O$-bearing atmosphere and oceans (after cooling). Shortly thereafter, Ahrens and colleagues experimentally studied impact outgassing of carbonaceous chondrites, carbonates, and hydrous minerals. They found that the volatiles released were composed primarily of $H_2O$ and $CO_2$ (e.g, see Lange and Ahrens, 1982a,b, 1986; Tyburczy and Ahrens, 1985, 1987, 1988; Tyburczy et al., 1986a,b). Due in large part to the work of Ahrens and colleagues, it is widely believed that conversion of at least some of the kinetic energy of accreted planetesimals into heat led to impact outgassing of volatiles during the accretion of the Earth and other terrestrial planets. Because $H_2O$ and $CO_2$ were the major species released by impact outgassing of the materials they studied (i.e., carbonaceous chondrites, carbonates, and serpentine) an atmosphere of impact origin has historically been called a steam atmosphere (e.g. Abe and Matsui, 1985; Lange and Ahrens, 1982a). Models of Earth's steam atmosphere and a magma ocean were developed by Abe and Matsui in a series of pioneering papers (Abe and Matsui 1985, 1987, Matsui and Abe 1986).

In this paper we use chemical equilibrium and chemical kinetic calculations to model chemistry of the atmospheres produced by impact outgassing during accretion of the Earth and other terrestrial planets. We explicitly assume that volatiles are outgassed during accretion and that the extent of outgassing and the speciation of volatiles are given by chemical equilibrium calculations. (We later examine these assumptions in section 4.3.) Throughout this paper we refer to the atmospheres so produced as "impact generated atmospheres" or in analogous terms. We are interested in answering the following questions. First, what is the chemistry of impact generated atmospheres produced from different types of chondritic material under the same set of nominal conditions (taken as 1500 K and 100 bars based on the work of Abe and Matsui 1987)? Second, what types of gases and volatile-bearing condensates are formed as a function of temperature and pressure over wide ranges? Third, is chemical equilibrium a good assumption for modeling chemistry of impact generated atmospheres? Fourth, what other elements are present in the impact generated atmospheres? Fifth, how does the presence of a magma ocean affect the atmospheric chemistry? Finally, what are the implications of our results for the chemistry of atmospheres formed during accretion of the Earth and other terrestrial planets?

Our paper is organized as follows. In Sections 2 and 3, we discuss our computational methods and results. In Section 4, we discuss previous modeling of impact degassed atmospheres, comparison with experimental results, kinetics, and application to magma ocean models. In Section 5, we summarize our results and answer the questions posed above. Preliminary results of our work are described in three abstracts (Fegley and Schaefer 2007, 2008, Schaefer and Fegley 2008).

## 2. Methods

We modeled the chemistry of impact produced atmospheres using thermochemical equilibrium calculations that were done with a Gibbs energy minimization code of the type described by van Zeggern and Storey (1970). Our calculations included thermodynamic data for ~930 condensed and gaseous compounds of 20 major rock-forming and volatile elements (Al, C, Ca, Cl, Co, Cr, F, Fe, H, K, Mg, Mn, N, Na, Ni, O, P, S, Si, Ti). We used the same database as in Schaefer and Fegley (2007). Calculations were performed for a wide range of pressures ($\log_{10} P$ of $-4$ to $+4$ bars) and temperatures (300 to 2500 K). Our nominal model has a pressure of 100 bars and a temperature of 1500 K based on the work of Abe and Matsui (1987). However, their estimate of pressure and temperature depends on the atmospheric composition, which they took as $H_2O$ plus $CO_2$. Impact produced atmospheres with different compositions may have different temperatures and pressures, which is why we have considered a wide $P - T$ range.

We consider atmospheres generated by impact outgassing of planetesimals made up of the major types of chondritic meteorite material – carbonaceous (CI, CM, CV), ordinary (H, L, LL), and enstatite (EH, EL) chondrites. We focused on chondritic material for several reasons (Schaefer and Fegley 2007), which we reiterate here. The chondrites are undifferentiated (i.e., unmelted) stony meteorites containing metal, silicate and sulfide, and are more abundant than other types of meteorites (achondrites, irons and stony irons). Chondritic material formed in the solar nebula and is believed to have been the material accreted by the Earth and other terrestrial planets during their formation. (e.g., Lewis and Prinn, 1984; Larimer 1971, Wänke 1981, Hart and Zindler 1986, Lodders and Fegley 1997; Lodders, 2000). For example, the oxygen isotope mixing model predicts that Earth incorporated 70% EH, 21% H, 5% CV, and 4% CI chondritic material (Lodders 2000). We use the major chondritic groups in our models because their volatile contents are well known, whereas the volatile contents of minor chondritic groups (e.g., the Rumurutiite (R) chondrites) are less well characterized or unknown. Similarly, we do not include achondritic material, which may also have contributed to the accretion of the terrestrial planets, because achondrites are fairly volatile-poor, and their volatile contents are poorly known. The minor chondrite groups and achondrites could be included into models of impact outgassing once their average volatile contents (C, H, N, S) are known. The compositions of the chondrites used here are from the following sources: (1) CI chondrites (Orgueil) – Lodders (2003), (2) average H, L, and LL chondrites – Schaefer and Fegley (2007), (3) CM chondrites (Murchison), CV chondrites (Allende), and average EH and EL chondrites – METBASE database, Koblitz (2005).

Our calculations assume that the impacting planetesimals and proto-planet have the same compositions, e.g. H chondritic material impacting an H chondritic proto-planet. This is equivalent to assuming that the composition of material accreted by the Earth did not change with time. This approach lays the foundation for studying more complicated and model dependent scenarios such as oxidized carbonaceous chondritic material impacting the Earth late in its formation (e.g., see Wänke 1981). We discuss chondrite mixing models in section 4.6.

Finally, these calculations do not model chemistry as a function of time in the vapor plume of an impacting planetesimal. Pressures and temperatures at the point of impact may reach very high values, but as the vapor plume expands the pressure and temperature drop. The expansion and cooling of vapor plumes depend upon the mass, velocity, impact angle, and composition of the impacting planetesimal. Quenching of chemistry within the vapor plume occurs when the cooling time of the vapor plume ($t_{cool}$) equals the characteristic reaction time ($t_{chem}$). Estimated quench temperatures for chemistry in impact vapor plumes range from 1000 to 2000 K and depend upon the reactions involved (e.g., see Fegley et al. 1986). Our results, which we discuss next, simply give the predicted chemical equilibrium composition as a function of pressure and temperature in an impact generated atmosphere on the early Earth or another rocky planet. In section 4.3 we consider kinetics and show that chemical equilibrium is a good assumption for our modeling.

**3. Results**

*3.1.1. Nominal Model: Bulk Gas Chemistry*

Figure 1 shows the predicted composition of an impact generated atmosphere having a total pressure of 100 bars as a function of temperature for CI, CV, H, EH, and EL chondritic material. Results for CM chondritic material are very similar to those for CI chondritic material and are not shown. Also, results for L and LL chondritic material are very similar to those for H chondritic material and are not shown. Minor gases, which contain the more volatile rock-forming elements and only become abundant at very high temperatures, are not plotted in Figure 1 for the sake of clarity. These minor gases include the most abundant gases formed by Cl (HCl, NaCl, KCl), F (HF, NaF), K (KOH, KCl, K), Na (NaOH, Na, NaCl), and P (PO, $PO_2$, $P_4O_6$). We discuss the chemistry of these rock-forming elements in section 4.5. Table 1 gives the major gas composition at our nominal conditions of 100 bars, 1500 K.

Figure 1 shows that $N_2$ is the major gas at low temperatures for all the impact generated atmospheres. As temperature increases, carbon and hydrogen are released into the gas phase and $N_2$ drops in abundance. However, $N_2$ is always the major N-bearing gas and $NH_3$ is generally much less abundant than $N_2$ at all temperatures in atmospheres formed from carbonaceous chondritic material. Water and $CO_2$ are the major gases released from carbonaceous chondritic material (CI, CM and CV). Steam is more abundant than $CO_2$ for CI and CM chondritic material, while $CO_2$ is more abundant than $H_2O$ for CV chondritic material. Carbon monoxide becomes more abundant with increasing temperature for CI, CM, and CV material (Fig 1a, 1b). Carbon and hydrogen are released at lower temperatures from CI and CM chondritic material than from CV chondritic material. This is probably due to thermal decomposition at low temperatures of the carbonates and hydrous minerals in CI and CM chondritic material. The $CH_4$ abundance peaks in the 500 – 100 K range, but it is not the major C-bearing gas in atmospheres formed by impact outgassing of carbonaceous chondrites. At 100 bars, $H_2$ is a minor gas and is always less abundant than water. Hydrogen sulfide is the major S-bearing gas at low temperatures but $SO_2$ becomes more important at higher temperatures.

Figure 1c shows major gases in atmospheres generated from ordinary chondritic material, exemplified by H chondritic material. Ordinary chondritic material is more reduced than carbonaceous chondritic material, which leads to production of a more reducing atmosphere. Molecular nitrogen is the major gas at low temperatures and is generally the major N-bearing gas except for a small range from 500 – 600 K where $NH_3$ becomes more abundant. This region is also shown as the $NH_3$ wedge in Figure 6a discussed later. Carbon is released primarily as $CH_4$, with CO becoming the major C-bearing gas at higher temperatures (1100 – 1200 K). Hydrogen is mainly released as $H_2$ with $H_2O$ being less abundant. Sulfur is primarily released as $H_2S$ and $SO_2$ is significantly less important.

The chemistry of atmospheres generated by impact outgassing of enstatite chondritic material is shown in Figures 1d and 1e. The results for EH and H chondritic material are similar. The primary difference is that outgassing of EH chondritic material yields more $N_2$ than $NH_3$ at all temperatures (cf. Figures 1c and 1d). Once again, carbon is released primarily as $CH_4$ at low temperatures with CO becoming dominant at higher temperatures (1200 – 1300 K). Overall, the atmospheres generated from H, EH, and EL chondritic material are similar, although CO is more important and $H_2$ is less important at the nominal pressure and temperature (see Table 1).

*3.1.2. Carbon Chemistry*

Figure 2 illustrates the major C-bearing species at chemical equilibrium as a function of temperature and pressure for outgassing of CI, CM, and CV chondritic material. The fields labeled CO, $CO_2$, graphite, $MgCO_3$, and $CaMg(CO_3)_2$ show the P – T regions where these compounds are the major form of carbon. The solid curves in Figure 2 are equimolar abundance lines. For example, the CO – $CO_2$ line in Figure 2a shows where CO and $CO_2$ have equimolar abundances in the atmosphere generated from CI chondritic material. Likewise the $CO_2$ – graphite line shows where $CO_2$ gas and graphite have the same molar abundances. The $CO_2$ – $CH_4$ dotted line in Figure 2c is also an equimolar abundance line, but in this case graphite is more abundant than either gas. At low temperatures, i.e., the lower parts of Figures 2a – 2c, impact outgassing is ineffective and most carbon condenses out as $MgCO_3$ (magnesite), graphite, or $CaMg(CO_3)_2$ (dolomite). The graphite field is very broad for CV material and covers almost the entire range of pressures below ~ 500 K. In general, $CO_2$ is the major C-bearing gas produced by outgassing of carbonaceous chondritic material over a wide range of temperatures and pressures. High temperatures and low pressures favor CO over $CO_2$, but small $CO_2$ fields exist at high T and low P in Figures 2b and 2c. Methane is never the major C-bearing gas in atmospheres generated from carbonaceous chondritic material.

In contrast, $CH_4$ is the major C-bearing gas from outgassing of ordinary and EH chondritic material. This is shown in Figure 3, which is interpreted the same way as explained above for Figure 2. Figure 3a shows that $CH_4$ or CO is the major C-bearing gas produced by outgassing ordinary chondritic material. Carbon dioxide is never the major C-bearing gas, and is always less abundant than $CH_4$ or CO. The graphite wedge in Figure 3a shows that most carbon condenses as graphite in this P – T region. The dotted line inside this region is the equimolar abundance line for $CH_4$ and CO, which has almost the same position in Figures 3a – 3c for H, EH, and EL chondritic material. Figure 3b shows that outgassing of EH chondritic material results in either CO or $CH_4$ being dominant with $CO_2$ always being less abundant. However, in this case the graphite field is larger and occurs at both low T and high P (bottom right) and also at intermediate pressures and temperatures (in the wedge). Figure 3c shows that outgassing of EL chondritic material is inefficient over a wide P – T range and mainly leads to condensation of carbon as graphite or $Fe_3C$ (cohenite) unless outgassing occurs at sufficiently high temperatures where CO is the major gas. Methane is the major C-bearing gas throughout most of the graphite field, while $CO_2$ is never the major carbon gas and is always less abundant than $CH_4$ or CO.

*3.1.3. Hydrogen Chemistry*

Figures 4 and 5 for hydrogen-bearing compounds are analogous to Figures 2 and 3 for carbon compounds. Figure 4 shows the major H-bearing species in atmospheres from outgassing of carbonaceous chondritic material. Figure 4a shows CI and CM chondritic material, for which the results are nearly identical. Outgassing at low temperatures and high pressures produces mainly liquid water (bottom right region of Figure 4a). Liquid water exists in equilibrium with $H_2O$ vapor, which has an abundance given by the saturated vapor pressure at the ambient temperature. Hydrous minerals, which are less abundant than liquid water, also exist in the liquid water field. The dashed and dotted lines show where liquid water and steam have equimolar abundances. Steam is the major H-bearing species at higher temperatures and $H_2$ is less abundant.

Figure 4b shows the major H-bearing species from outgassing of CV chondritic material. In this case, most hydrogen is in glaucophane, tremolite, or anthophyllite at low to intermediate temperatures over a wide pressure range. This occurs because as the impact-degassed atmosphere cools, significant amounts of water are trapped in these hydrous silicate minerals. Steam is the major hydrogen gas at higher temperatures, and it dissociates to atomic H and OH at the lowest pressures and highest temperatures shown. Molecular $H_2$ is always less abundant than $H_2O$ in atmospheres produced from CV chondritic material.

Figure 5 shows the major H-bearing species in atmospheres from outgassing of ordinary chondritic and enstatite chondritic material. At low temperatures and high pressures, most of the hydrogen condenses into talc or glaucophane and outgassing is inefficient. At lower pressures and higher temperatures, $CH_4$ becomes the dominant H-bearing species. At still higher temperatures, the abundance of $CH_4$ drops (e.g., see Fig. 1c), and $H_2$ becomes the dominant H-bearing species. Only at very high temperatures and low pressures, does steam (and then H gas) become the most abundant H-bearing species. Otherwise $H_2O$ is less abundant than other H-bearing gases in the atmospheres generated by outgassing ordinary chondritic material. The equal abundance lines in this figure take into account the different number of H atoms in talc, $CH_4$, $H_2$, and H. Thus for example, the $CH_4 - H_2$ line is where the molar abundance of H atoms in the two gases is equal because one mole of $CH_4$ contains 4 H atoms while one mole of $H_2$ contains 2 H atoms.

Figures 5b and 5c show the major H-bearing species in atmospheres generated by outgassing EH and EL chondritic material. The results for enstatite chondritic material are similar to those for H chondritic material. At low temperatures and high pressures, outgassing is inefficient and most hydrogen is in a hydrous silicate (talc for EH chondrites, and glaucophane for EL chondrites). As temperature increases, outgassing is more efficient and $CH_4$ becomes the dominant H-bearing species. At higher temperatures, $H_2$ becomes more abundant than $CH_4$, and at even higher temperatures, H (g) becomes the most abundant species. The equal abundance lines in Figure 5 are interpreted the same way as those in Figure 4.

*3.1.4. Nitrogen Chemistry*

At all pressures and temperatures, $N_2$ is the major nitrogen-bearing species for atmospheres degassed from CI, CM, and CV chondritic material. However, as shown in Figure 6, $N_2$ is almost, but not quite always the major N-bearing species in atmospheres degassed from ordinary and enstatite chondritic material. Ammonia is the major N-bearing gas at high pressures and intermediate temperature (shown by the wedge-shaped region in Figure 6a and the $NH_3$ peak in Figure 1c). The results for EH and EL enstatite chondritic material are nearly identical to one another. In these cases, the major N-bearing species is typically $N_2$ (g). However, at high pressure and low temperature, nitrogen condenses into solid iron nitride $Fe_4N$. The equal abundance lines in Figure 6 are analogous to those in Figure 5 and are calculated in terms of equal abundances of N atoms in each species.

**4. Discussion**

*4.1. Comparison to other calculations*

There are very few calculations to which we can compare our results because the outgassing of chondritic material has been modeled theoretically by only a few people. Bukvic (1979) did chemical equilibrium models of outgassing from the outer layers of an Earth-like planet. He used H chondritic and various mixtures of H and CI chondritic compositions, and an Earth-like geotherm. In all cases he found that the volcanically outgassed volatiles were dominated by $CH_4$ and $H_2$. This study, which was suggested by J. S. Lewis, was briefly discussed by him in Lewis and Prinn (1984).

Schaefer and Fegley (2007) did chemical equilibrium models of thermal outgassing of H, L, and LL ordinary chondritic material along model P – T profiles for the asteroid 6 Hebe, the Earth's geotherm, and also as a function of temperature (400 – 2000 K) and pressure ($10^{-4}$ to $10^4$ bars). They applied their results to volatile release during thermal metamorphism of ordinary chondritic material and the Earth's early atmosphere. They found that $CH_4$ and $H_2$ are the major volatiles outgassed from ordinary chondritic material under a wide variety of conditions. However, Schaefer and Fegley (2007) did not present results for the nominal pressure and temperature conditions proposed for Earth's putative steam atmosphere or consider several of the questions addressed in this paper.

Hashizume and Sugiura (1998) did chemical equilibrium calculations of carbon speciation during metamorphism of ordinary chondritic material. They reported that CO was the major C-bearing gas, but as noted by Schaefer and Fegley (2007), after reading a preprint of their paper Professor Sugiura discovered that Hashizume and Sugiura (1998) used incorrect Gibbs energy data for $CH_4$. When this mistake was corrected he found that $CH_4$ was the major C-bearing species produced during metamorphism of ordinary chondritic material, in agreement with the results of Schaefer and Fegley (2007) and those given in Figure 3a here.

Hashimoto and Abe (1995) and Hashimoto et al. (2007) performed outgassing calculations for an average CI composition. However, they used a small number of elements in their calculations (H, C, N, O, S, and Fe), and did not actually consider the other elements (e.g., Ca, Al, Si, Ti, P, etc.) found in CI chondrites. Nevertheless, the results of their calculations appear to be broadly similar to ours with a few exceptions. Their results (Fig. 1 and 2 in Hashimoto et al. 2007) do not include the abundance of $H_2O$ (g), which our calculations show to be the most abundant gas formed by heating CI chondritic material. It is not entirely clear whether they included $H_2O$ vapor in their calculation of the gas-phase equilibria, although the similarity of their results to ours suggests that they did. However, the presentation of their results gives the misleading impression that $H_2$ and CO are the two most abundant gases at high temperature, whereas our calculations show that they account for less than 25% of the total atmosphere at 2500 K.

Additionally, their calculations use incorrect data for several sulfur species. Hashimoto et al. (2007) calculate incorrectly that the major sulfur-bearing gases at high temperatures are $S_2O$ and SH. They took the thermodynamic data for these species from the JANAF tables (Chase, 1998). However, as discussed by Lodders (2004), the data for both $S_2O$ (g) and SH (g) in the JANAF tables are incorrect. We used the corrected data for these gases from Lodders (2004) in our calculations and find that the major S-bearing gas from CI chondritic material is $H_2S$ at temperatures below 2200 K and $SO_2$ at temperatures above 2200 K. Also, either $H_2S$ or $SO_2$ is the major S-bearing gas in terrestrial volcanic gases (e.g., Symonds et al. 1994).

*4.2. Impact experiments.*

Ahrens and colleagues have performed shock devolatilization experiments with serpentine ($Mg_3Si_2O_5(OH)_4$), calcite ($CaCO_3$), and the Murchison CM chondrite (e.g, Lange and Ahrens 1982a,b, 1986; Tyburczy and Ahrens 1985, Tyburczy et al. 1986a,b). These experiments are designed to simulate impact induced outgassing of carbonaceous chondritic material and find that $H_2O$ and $CO_2$ are the major gases released. This *qualitative* conclusion is in complete agreement with our calculations, which show that $H_2O$ and $CO_2$ are the major gases released from CI and CM carbonaceous chondritic material outgassed over a wide P – T range (e.g., Figures 1a, 1b; 2a, 2b; 4a, 4b; Table 1). Unfortunately we are unable to *quantitatively* compare our results to their experimental results because the composition of the devolatilized gas was analyzed in only a few experiments (e.g., for $H_2O$ and $H_2$). Mass-balance of the system was not maintained, and in some instances, the captured gases reacted with the experimental apparatus, which may have altered the speciation of the evolved volatiles. We suggest that similar experiments done with a wider range of chondritic materials and with more controlled gas analysis would be useful.

Russian scientists have also studied high-velocity impacts through a different technique: laser pulse heating. They used terrestrial rocks such as augite, basalt, peridotite and gabbro, as well as the Tsarev (L5), Kainsaz (CO), and Murchison (CM) chondrites (Gerasimov and Mukhin 1984, Mukhin et al. 1989). The released volatiles for all materials had broadly similar compositions, with the major gases including $H_2$, $N_2$, CO, $CO_2$, hydrocarbons, $SO_2$, $H_2S$, HCN, $CH_3CHO$, and water vapor. (While water vapor was detected, its abundance was not determined because of its strong adsorption on their equipment.) Samples were heated to temperatures of 3500 – 4500 K for ~$10^{-5}$ s, and the gas composition quenched at ~3000 K (Gerasimov et al. 1998). As we have already discussed above, quench temperatures for impacts depend on many variables but may range from ~1000 – 2000 K. Mukhin et al. (1989) and Gerasimov et al. (1998) suggest that gas compositions are only weakly dependent on temperature at the conditions of a high-velocity impact, and therefore the difference between the quench temperatures of their experiment versus actual quench temperatures found during an impact should make little difference. However, we have found that the gas composition dependence on temperature varies based on the impactor material. For instance, the composition of gases released from the CM chondrite are weakly dependent on temperature at very high temperatures: at 3000 K the major gases are $H_2O$, CO, $H_2$, and OH, and at 2000 K are $H_2O$, $H_2$, $CO_2$, and CO, in order of decreasing abundance. In contrast, for an L-chondrite, the major gases at 3000 K are Fe, SiO, Mg, and Na, and at 2000 K are $H_2$, CO, $H_2O$, and K vapor. The reduced material (L chondrite) has much larger temperature dependence than the oxidized material (Murchison). This should not be too surprising since the L chondrite has a smaller volatile content than the CM chondrite, so a greater portion of the impact energy will be used to vaporize silicate material.

Gerasimov et al. (1998) also found molecular and atomic oxygen with abundances up to ~30% of the released vapor. In our calculations, this corresponds to a temperature range of ~3500 – 4000 K, which agrees well with Gerasimov's estimated temperatures. At such temperatures, silicate material vaporizes. As temperature cools, the free oxygen recombines with the silicate vapors and forms condensates. The researchers found that condensation of silicate material trapped significant quantities of volatiles such as $H_2O$, $CO_2$ and N in the condensed phase, up to 10 wt% water, 4 wt% $CO_2$ and 0.1 wt% N (Gerasimov et al., 1994a,b, 1999). This effect may significantly reduce the amount of atmosphere that would be produced by a planetesimal impact; however, as the experiments had both shorter quenching times and higher temperatures than large impacts, it is uncertain whether the same degree of trapping would occur. Furthermore, it has been shown by Kress and McKay (2004) and Sekine et al. (2003) that silicate and metal particles in the atmosphere may actual push the system closer to equilibrium by catalyzing gas phase reactions.

*4.3. Kinetics*

Table 1 shows that at the nominal conditions of 1500 K and 100 bars, the four major gases in the impact generated atmospheres are $H_2$, $H_2O$, $CO_2$ and CO. These results are based on chemical equilibrium calculations and thus we now consider whether or not gas phase reactions are rapid enough for equilibrium to be achieved at the nominal conditions.

These four gases are equilibrated by the water gas reaction which is

$$H_2O + CO = H_2 + CO_2 \qquad (1)$$

Reaction (1) proceeds via elementary reactions such as

$$CO + OH \rightarrow H + CO_2 \qquad (2)$$

$$\log_{10} k_2 = 3.94 \times 10^{-4} T - 12.95 \text{ cm}^3 \text{ molecule}^{-1} \text{ s}^{-1} \qquad (3)$$

$$H + H_2O \rightarrow H_2 + OH \qquad (4)$$

$$k_4 = 4.11 \times 10^{12} \left(\frac{T}{298}\right)^{1.6} \exp(-9720/T) \text{ cm}^3 \text{ s}^{-1} \qquad (5)$$

The rate constants for reactions (3) and (4) are from Baulch et al. (1976) and Baulch et al (1992). The chemical lifetimes ($t_{chem}$) for conversion of CO into $CO_2$ and of $H_2O$ into $H_2$ are given by

$$t_{chem}(CO) = \frac{1}{k_2[OH]} \text{ s} \qquad (6)$$

$$t_{chem}(H_2O) = \frac{1}{k_4[H]} \text{ s} \qquad (7)$$

where [OH] and [H] are the number densities of OH radicals and H atoms, and $k_2$ and $k_4$ are the rate constants for reactions (2) and (4), respectively.

At 1500 K and 100 bars, the nominal conditions based on the work of Abe and Matsui (1987), the chemical lifetime of CO is 0.03 – 0.4 seconds in impact generated atmospheres that are produced from carbonaceous (CI, CM, CV) or ordinary (H) and enstatite (EH, EL) chondritic material, respectively. The chemical lifetime for $H_2O$ conversion into $H_2$ is virtually instantaneous ($10^{-25}$ seconds) for atmospheres generated from all types of chondritic material studied. The reverse of reaction (2) and the reverse of reaction (4) are similarly fast. Thus, the water gas reaction reaches equilibrium at the nominal conditions and all four gases are equilibrated with one another.

A related question is the lowest temperature at which reaction (1) attains equilibrium. Figure 7 answers this question and shows the chemical lifetimes for CO conversion into $CO_2$ as a function of temperature. Only three curves for CI, CV, and H chondritic material are shown because the other curves are virtually identical to the ones shown (EH and EL are like H and CM is like CI). The dashed cooling line is the radiative cooling time for large impacts calculated from the Stefan-Boltzmann equation as done by Fegley et al. (1986). The chemical lifetimes for CO conversion to $CO_2$ are shorter than the radiative cooling time down to ~ 950 – 860 K for H or CV and CI chondritic material, respectively. The chemical lifetime for reaction (4) is always much shorter than plausible cooling times. Thus, our results show that the water gas reaction should reach equilibrium in the impact generated atmospheres down to 950 – 860 K.

Figure 1 shows that $CH_4$ becomes increasingly important at lower temperatures. Thus, a third important question is whether or not chemical equilibrium is maintained to temperatures where $CH_4$ is an important C-bearing gas. Carbon monoxide or $CO_2$ are converted into $CH_4$ via net thermochemical reactions such as

$$CO + 3H_2 = CH_4 + H_2O \tag{8}$$

$$CO_2 + 4H_2 = CH_4 + 2H_2O \tag{9}$$

The rate determining step that is responsible for converting CO or $CO_2$ into $CH_4$ is still debated in the literature (e.g., see the brief review in Visscher and Fegley 2005), but is either

$$H_2 + H_2CO \rightarrow CH_3 + OH \tag{10}$$

$$k_{10} = 2.3 \times 10^{-10} \exp(-36,200/T) \text{ cm}^3 \text{ s}^{-1} \quad (11)$$

$$\text{H} + \text{H}_2\text{CO} \rightarrow \text{CH}_3\text{O} + \text{H} \quad (12)$$

$$k_{12} = 1.5 \times 10^{11} T \exp(-12,880/T) \text{ s}^{-1} \quad (13)$$

Thus, the chemical lifetime for CO reduction to $CH_4$ is either

$$t_{chem}(CO) = \frac{[CO]}{k_{10}[H_2][H_2CO]} \text{ s} \quad (14)$$

$$t_{chem}(CO) = \frac{[CO]}{k_{12}[H][H_2CO]} \text{ s} \quad (15)$$

Likewise, the chemical lifetime for $CO_2$ reduction to $CH_4$ is either

$$t_{chem}(CO_2) = \frac{[CO_2]}{k_{10}[H_2][H_2CO]} \text{ s} \quad (16)$$

$$t_{chem}(CO_2) = \frac{[CO_2]}{k_{12}[H][H_2CO]} \text{ s} \quad (17)$$

The two different rate determining steps give dramatically different chemical lifetimes for reduction of CO and $CO_2$ to $CH_4$. If reaction (10) is the rate determining step formation of $CH_4$ should stop at high temperatures of ~ 1820 K or 1380 K for CI (and CM) or H (and EH, EL) chondritic material, respectively. Also, $CH_4$ formation takes significantly longer than the radiative cooling time in impact generated atmospheres from CV chondritic material. Figure 1 shows that the 1820 K quench temperature is much higher than the temperature range of 500 – 600 K where $CH_4$ becomes important in atmospheres formed from outgassing CI and CM chondritic material. However, the $CH_4$ – CO equal abundance point is at 1250 K, which is only slightly less than the 1380 K quench temperature for atmospheres formed from H, EH, and EL chondritic material (e.g., see Figures 1 and 3). On the other hand, if reaction (12) is the rate determining step, our results show that $CH_4$ formation remains favorable to temperatures below 800 K. It is unclear whether reaction (10) or (12) is the rate determining step and arguments have been made both ways (cf. Visscher and Fegley 2005).

Although gas phase reactions may become quenched at lower temperatures, experimental data and observations of terrestrial volcanic gases suggest equilibrium is reached down to lower temperatures where $CH_4$ becomes dominant. For example, analyses of volcanic gases show that chemical equilibrium is maintained to temperatures as low as ~700 K (e.g. Gerlach 1993, Symonds et al. 1994). Reaction (9) has been studied experimentally down to ~ 520 K (e.g., Randall and Gerard 1928) and equilibrium was reached with aid of a metal catalyst. Theoretical models of iron meteorite (Sekine et al. 2003) and cometary (Kress & McKay 2004) impacts suggest that $CH_4$ formation from $H_2$ and CO continues down to low temperatures due to catalysis by solid grains. Thus we conclude that reactions (8) and (9) may also equilibrate down to low temperatures, possibly in the gas phase if reaction (12) is rate limiting, and probably with catalysis by solid grains.

Large impacts may also create favorable conditions in which equilibrium can be achieved. Laboratory studies have shown that shocked serpentine re-equilibrates with the gas phase orders of magnitude more quickly than unshocked serpentine, which suggests that surface-atmosphere equilibrium during accretion was likely rapid (Tyburczy and Ahrens 1987, 1988).

We also considered kinetic inhibition of condensate formation at low temperatures. For example, graphite, magnesite, and dolomite form at chemical equilibrium at low temperatures and become the major C-bearing phases (see Figures 2 and 3). What happens if these solids do not precipitate out of the gas? Carbon dioxide remains the major C-bearing gas in the atmospheres formed from CI and CM chondritic material and $CH_4$ is the second most abundant carbon gas. Carbon chemistry in atmospheres produced from CV and EH chondritic material is

- 21 -

significantly different if graphite does not form. Coronene ($C_{24}H_{12}$), which is a polyaromatic hydrocarbon with seven benzene rings, becomes the major C-bearing gas and $CH_4$ is the second most abundant carbon gas. Methane is the major carbon gas at low temperatures if graphite does not condense from atmospheres produced from EL chondritic material. Likewise, water vapor and liquid water are the major H-bearing species in the absence of hydrous silicate formation.

*4.4. Magma Ocean Models*

The presence of a magma ocean at the planet's surface has implications for atmospheric chemistry, primarily for steam. Water vapor is very soluble in magma, much more so than the other volatiles in the outgassed atmospheres. The presence of a magma ocean would reduce the amount of steam, and likely $CO_2$, present in the atmosphere. The atmospheres produced from carbonaceous chondritic material, which produce significantly more steam than the other types of chondritic material, would be affected the most by the presence of a magma ocean. Moore et al. (1998) modeled the solubility of $H_2O$ in magmas up to 3 kilobars. Using their model, we estimate that at 1200 K and 1 kilobar, the amount of $H_2O$ dissolved in the magma ocean is ~2% for H and EH chondritic atmospheres, up to ~10% for CI chondritic atmospheres. Mass balance calculations assuming a 10 km deep global magma ocean show that dissolution of water in the magma ocean would change the atmospheric abundance of water vapor very little for the atmospheres from H and EH chondritic material, but may reduce the water vapor abundance by almost 50% in an atmosphere from CI chondritic material.

Our models predict that an impact-generated atmosphere may be significantly different than the pure steam or $H_2O - CO_2$ atmospheres used in models involving magma oceans sustained through greenhouse warming (e.g., Abe and Matsui, 1985; Abe, 1997; Matsui and Abe, 1986). Steam atmospheres produced from carbonaceous chondritic material contain significant amounts of $CO_2$, CO, and $H_2$. Smaller amounts of $H_2S$ and $SO_2$ are also predicted. Carbon monoxide, $H_2S$, and $SO_2$ have IR bands of varying strength but $H_2$ does not. The atmospheres produced from H, L, LL, and EH chondritic material are dominantly $H_2$ and that from EL chondritic material is dominantly CO at the nominal conditions. Minor or trace amounts of various greenhouse gases are present (see Table 1). The greenhouse warming produced by the $H_2$- or CO-rich atmospheres should be significantly less than that in the $H_2O$-rich atmospheres modeled in the literature. Consequently, surface conditions may be cooler and lower pressure than those based on the work of Abe and Matsui. This is one reason why we have considered a wide range of P, T conditions in our modeling. Modeling of the expected surface conditions for the $H_2$-, CO-rich atmospheres predicted by our calculations is beyond the scope of this work.

*4.5. Rock-Forming Elements in the Gas*

Table 1 shows that several gases, which are grouped together as "other", comprise 0.17% (CM) to 1.02% (CV) of the atmospheres generated from the different types of chondritic material. These gases contain rock-forming elements such as F, Cl, K, Na, P, and S. We now discuss the gas phase chemistry of these rock-forming elements at the nominal conditions (1500 K, 100 bars) shown in Table 1. Hydrogen chloride is the major Cl-bearing gas in all cases. The percentage of total chlorine in the gas varies from 95% (CI) to 49% (H) to 34-38% (EL, EH), and to 10% (CV) for the different atmospheres. The HCl abundance varies from 0.44% (EH) to 0.06% (CM) and averages 0.20%. The remaining chlorine that is not in HCl condenses out of the impact generated atmospheres as chlorapatite, sodalite, and alkali halides.

Hydrogen fluoride is the major F-bearing gas and has an abundance of 10 parts per million by volume (0.001 %). In all cases, most fluorine condenses out as fluorapatite and very

- 23 -

little of the total fluorine present is in the impact produced atmospheres, e.g., 0.02% for EL chondritic material to a maximum of 4.3% for CI chondritic material.

The alkali chlorides (KCl, NaCl) are the major gases of the alkali metals in all cases and they generally have equal abundances of 0.01 – 0.08%, although KCl reaches 0.17% in atmospheres generated from LL chondrites. The dimers $K_2Cl_2$, $Na_2Cl_2$ are the second most abundant gases for K and Na, but are much less abundant than KCl or NaCl. Most of all K and most of all Na condense out as feldspar, micas, and alkali halides and only 0.2 – 4.6% of Na and 0.7 – 11% of K are in the gas at 1500 K, 100 bars.

Sulfur chemistry was discussed earlier in section 3. Hydrogen sulfide is generally the major sulfur gas although $SO_2$ also becomes important. The minor sulfur gases OCS, $S_2$, SO and $S_2O$ are included in the "other" category in Table 1. Sulfur is mainly condensed out as sulfides and only 0.1 – 18% of total sulfur is in the gas phase.

Phosphorus is the last rock-forming element of any consequence in the gas phase. Its gas phase chemistry is complex and is dominated by oxides (PO, $PO_2$, $P_4O_6$). Phosphorus is more volatile in atmospheres formed from the more reducing H, EH (90% P in the gas), and EL chondritic material and is less volatile in atmospheres formed from the more oxidizing CI, CM, and CV (~ 0% P in the gas) chondritic material.

*4.6. Mixtures of Chondritic Material*

The material accreted by the Earth and other terrestrial planets was probably a mixture of chondritic material (e.g., Barshay 1981, Hart and Zindler 1986, Lodders 2000, Lodders and Fegley 1997, Wänke 1981). However, as mentioned earlier, our calculations assume that the impacting planetesimals and proto-planet have the same compositions, e.g. H chondritic material impacting an H chondritic proto-planet. This was done because the results for mixtures of chondritic material can be derived by linear mixing of the results for the end-member compositions. For example, the oxygen isotope mixing model (OIM model) predicts the Earth formed from 70% EH, 21% H, 5% CV, and 4% CI chondritic material (Lodders 2000). A linear combination of the results from Table 1 gives an atmospheric composition of 41% $H_2$, 28% CO, 19% $H_2O$, 8% CO, with other gases making up the remaining 4%. Likewise, the OIM model predicts that Mars formed from 85% H, 11% CV, and 4% CI chondritic material (Lodders and Fegley 1997). Using data from Table 1, the predicted atmospheric composition is 53% $H_2$, 21% CO, 20% $H_2O$, 3% $CO_2$, with other gases comprising the remaining 3%.

We also used the results in Table 1 to estimate the amounts of CI or CM chondritic material necessary to produce atmospheres with at least 50% steam. These calculations show that most of the accreted material must be either CI or CM chondritic material, e.g. 62% CI and 38% H chondritic material or 57% CM and 43% H chondritic material. It is unlikely that most of the material accreted by the Earth was like CI or CM carbonaceous chondrites. However accretion of a late CI-like veneer, comprising a few percent of Earth's total mass, may be required to explain noble metal abundances in Earth's mantle (e.g., Wänke 1981).

## 5. Summary

We now return to the questions posed in the introduction and provide answers to them. First, Table 1 lists the chemistry of impact generated atmospheres derived from the major types of chondritic material at the nominal conditions of 1500 K, 100 bars. This table shows that steam atmospheres are produced from CI and CM chondritic material, in agreement with the results of impact outgassing experiments by Ahrens and colleagues. Outgassing of ordinary (H, L, LL) and enstatite (EH) chondritic material yields $H_2$-rich atmospheres, EL chondritic material gives a CO-rich atmosphere, and CV chondritic material produces a $CO_2$-rich atmosphere.

Second, calculations done over wide P – T ranges show that the results in Table 1 remain valid at other pressures and temperatures. Thus, $H_2O$, $CO_2$, and $N_2$ remain the major H-, C-, and N-bearing gases, respectively in atmospheres generated from carbonaceous (CI, CM, CV) chondritic material. The one exception (for CV material) is that $CH_4$ becomes the major carbon gas at low temperatures. Figures 3 and 5 show the situation is more complicated for atmospheres generated from ordinary and enstatite chondritic material because either CO or $CH_4$ and either $H_2$ or $CH_4$ are the major C-, and H-bearing gases, respectively, as a function of pressure and temperature. Also, $N_2$ is almost, but not quite always, the major N-bearing gas because $NH_3$ becomes dominant in a small P – T wedge for H chondritic material (Figure 6a).

Third, gas phase kinetic calculations show that equilibrium is reached in all cases at the nominal conditions (1500 K, 100 bars) and is also reached at all temperatures for equilibria between the four major gases $H_2$, $H_2O$, $CO_2$, and CO. Depending on the rate determining step for CO (or $CO_2$) reduction to $CH_4$, gas phase formation of $CH_4$ may not occur below 1380 K. However, experimental studies of $CO_2$ reduction to $CH_4$ show that equilibrium is reached down to ~ 520 K with aid of metal catalysts. Thus, $CH_4$ formation probably occurs down to low temperatures due to catalysis by solid grains.

Fourth, at high temperatures small amounts of F, Cl, K, Na, P, and S are also present in the impact generated atmospheres. The fraction of each element vaporized and its speciation in the gas depend upon the pressure, temperature, and type of chondritic material and are discussed in section 4.5. Fifth, the major effect of a magma ocean is dissolution of significant amounts of $H_2O$ from atmospheres generated by CI or CM chondritic material.

Finally, our results have several implications for atmospheric chemistry during accretion of the Earth and other terrestrial planets. The most important one is that $H_2O$-rich (steam) atmospheres only form if most of the material accreted is like CI or CM carbonaceous chondrites. For example, 57% (CM) to 62% (CI) of the accreted material in a binary mixture of carbonaceous and ordinary (H) chondritic material must be carbonaceous to have at least 50% steam in the outgassed atmosphere. Otherwise $H_2O$ is at best the second most abundant gas. Another implication is that the temperature and pressure conditions derived assuming $H_2O$ or $H_2O – CO_2$ atmospheres are probably valid only for carbonaceous chondritic planets because the greenhouse warming in $H_2$-, or CO-, or $CO_2$-rich atmospheres may be different than in pure steam or steam – $CO_2$ atmospheres. Finally, our results may be useful for interpreting and/or guiding spectroscopic observations of extrasolar terrestrial planets during their accretion.

## 6. Acknowledgments

We acknowledge support from the NASA Astrobiology and Origins Programs and interesting discussions with Tom Ahrens

## 7. References.

**Table 1.** Major Gas Compositions of Impact-generated Atmospheres from Chondritic Planetesimals at 1500 K and 100 bars

| Gas (vol %) | CI | CM | CV | H | L | LL | EH | EL |
|---|---|---|---|---|---|---|---|---|
| $H_2$ | 4.36 | 2.72 | 0.24 | **48.49** | **42.99** | **42.97** | **43.83** | 14.87 |
| $H_2O$ | **69.47** | **73.38** | 17.72 | 18.61 | 17.43 | 23.59 | 16.82 | 5.71 |
| $CH_4$ | $2\times10^{-7}$ | $2\times10^{-8}$ | $8\times10^{-11}$ | 0.74 | 0.66 | 0.39 | 0.71 | 0.17 |
| $CO_2$ | 19.39 | 18.66 | **70.54** | 3.98 | 5.08 | 5.51 | 4.66 | 9.91 |
| CO | 3.15 | 1.79 | 2.45 | 26.87 | 32.51 | 26.06 | 31.47 | **67.00** |
| $N_2$ | 0.82 | 0.57 | 0.01 | 0.37 | 0.33 | 0.29 | 1.31 | 1.85 |
| $NH_3$ | $5\times10^{-6}$ | $2\times10^{-6}$ | $8\times10^{-9}$ | 0.01 | 0.01 | $9\times10^{-5}$ | 0.02 | $5\times10^{-5}$ |
| $H_2S$ | 2.47 | 2.32 | 0.56 | 0.59 | 0.61 | 0.74 | 0.53 | 0.18 |
| $SO_2$ | 0.08 | 0.35 | 7.41 | $1\times10^{-8}$ | $1\times10^{-8}$ | $3\times10^{-8}$ | $1\times10^{-8}$ | $1\times10^{-8}$ |
| Other[a] | 0.25 | 0.17 | 1.02 | 0.33 | 0.35 | 0.41 | 0.64 | 0.29 |
| Total | 99.99 | 99.96 | 99.95 | 99.99 | 99.97 | 99.96 | 99.99 | 99.98 |

[a]Other includes gases of the rock-forming elements Cl, F, K, Na, P, and S. See text.

Figure Captions

Figure 1. Major gas compositions produced by impact outgassing of different types of chondritic planetesimals at the nominal pressure of 100 bars as a function of temperature from 300 to 2500 K. The panels show outgassing of (a) CI, (b) CV, (c) H, (d) EH, and (e) EL chondritic material.

Figure 2. Major carbon-bearing species in atmospheres formed from carbonaceous chondritic material. Fields indicate where each compound is the major C-bearing species. Lines indicate a 1:1 ratio of C in each species (e.g, $CO/CO_2 = 1$, $CO_2$/graphite $= 1$). (a) CI, (b) CM, (c) CV.

Figure 3. Major carbon-bearing species in atmospheres formed from ordinary (H) and enstatite (EH, EL) chondritic material. Fields indicate where each compound is the major C-bearing species. Lines indicate a 1:1 ratio of C in each species (e.g, $CO/CO_2 = 1$, $CO_2$/graphite $= 1$). (a) H, (b) EH, (c) EL.

Figure 4. Major hydrogen-bearing species in atmospheres formed from carbonaceous chondritic material. Fields indicate where each compound is the major H-bearing species. Lines indicate a 1:1 ratio of H in each species (a) CI, (b) CM, (c) CV.

Figure 5. Major hydrogen-bearing species in atmospheres formed from ordinary (H) and enstatite (EH, EL) chondritic material. Fields indicate where each compound is the major H-bearing species. Lines indicate a 1:1 ratio of H in each species. (a) H, (b) EH, (c) EL.

Figure 6. Major nitrogen-bearing species in atmospheres formed from ordinary (H) and enstatite (EH, EL) chondritic material. Fields indicate where each compound is the major N-bearing species. Lines indicate a 1:1 ratio of N in each species. (a) H, (b) EH and EL.

Figure 7. The chemical lifetime for CO conversion to $CO_2$ in impact generated atmospheres from H, CV, and CI chondritic material. The curves for EH and EL or CM chondritic material are very similar to the curves for H or CI chondritic material, respectively.

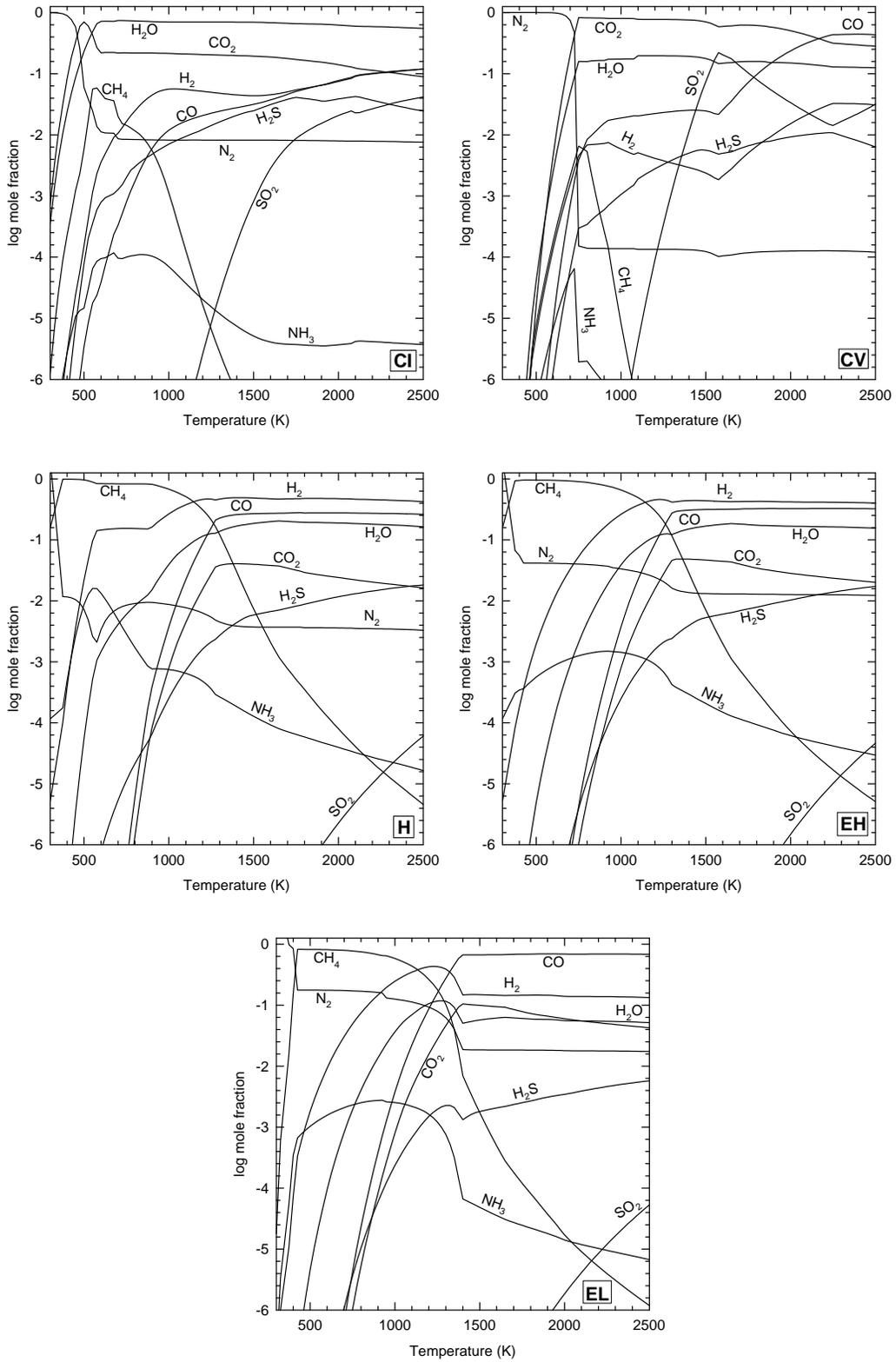

Figure 1. Major gas composition (a) CI, (b) CV, (c) H, (d) EH, (e) EL

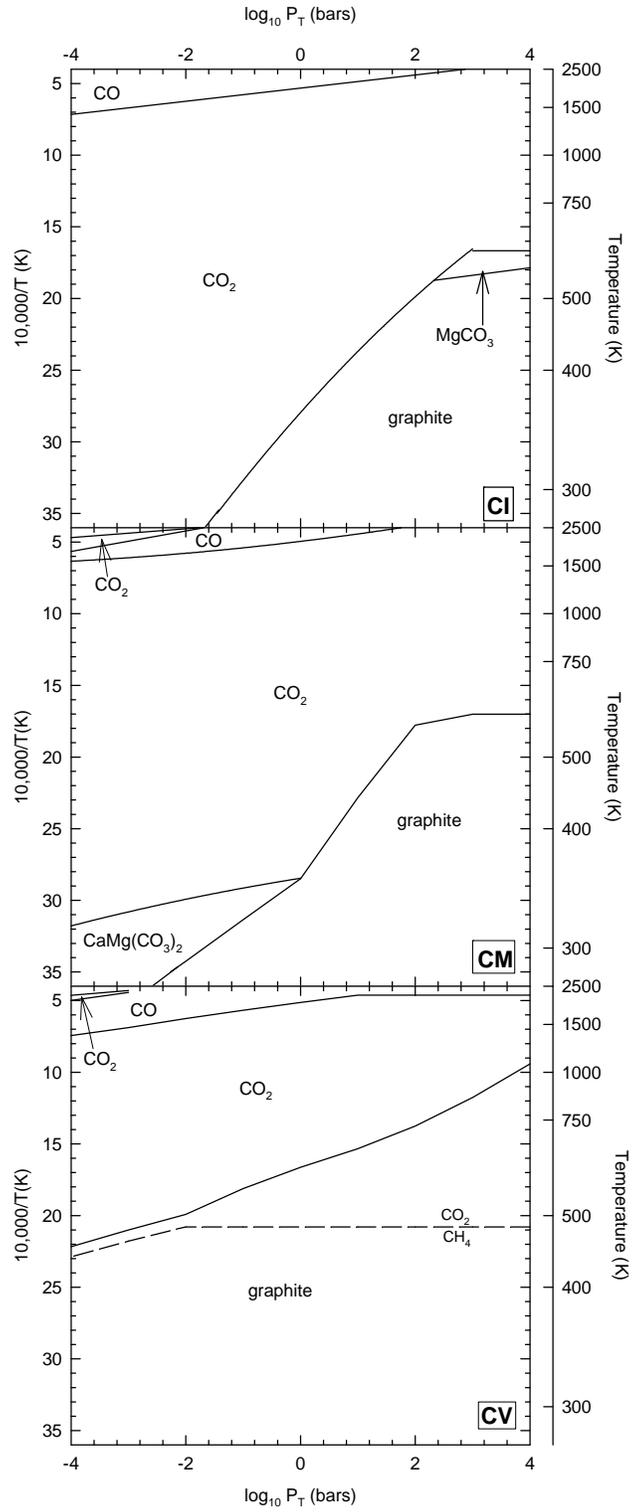

Figure 2. Carbon in (a) CI, (b) CM, (c) CV

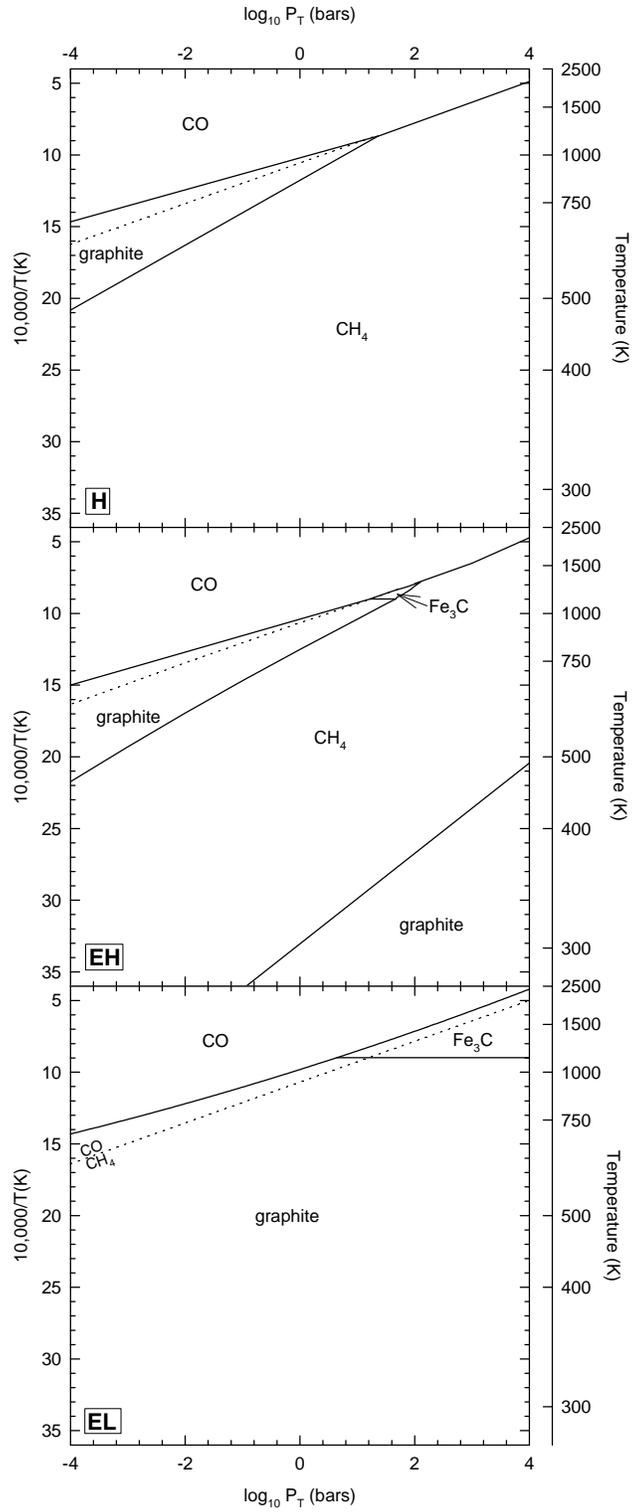

Figure 3. Carbon in (a) H, (b) EH, and (c) EL.

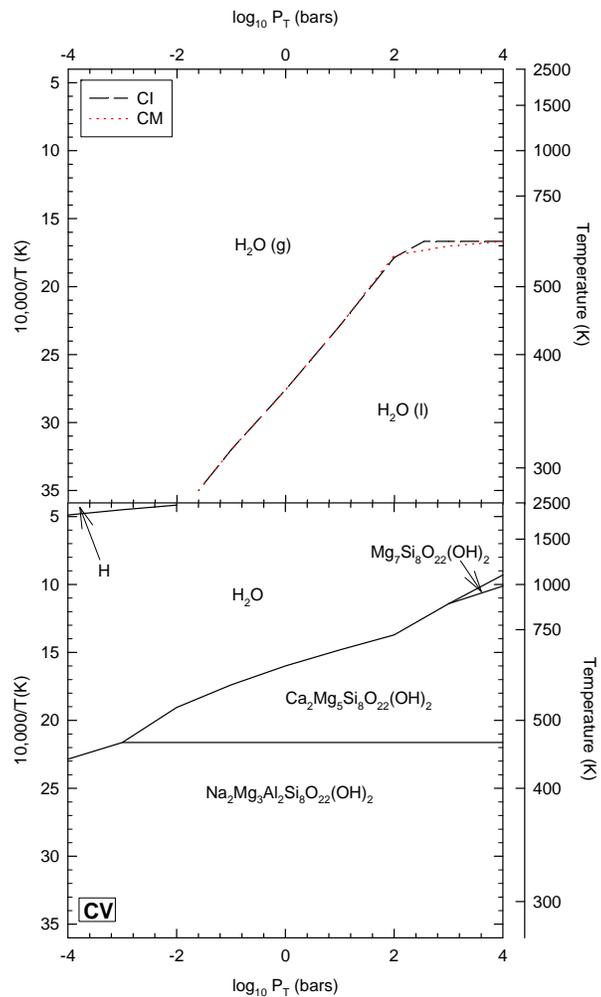

Figure 4: Hydrogen ratios in (a) CI, (b) CM, (c) CV

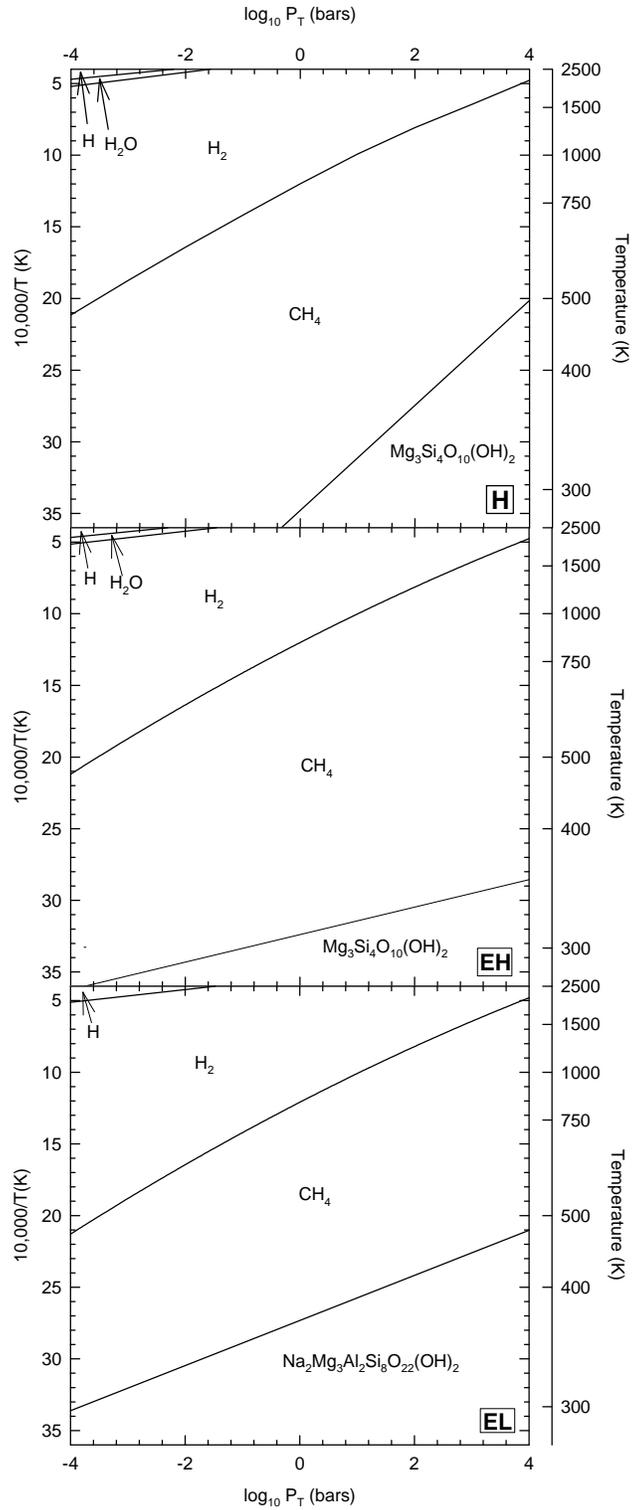

Figure 5: Hydrogen ratios in (a) H, (b) EH, (c) EL chondrites.

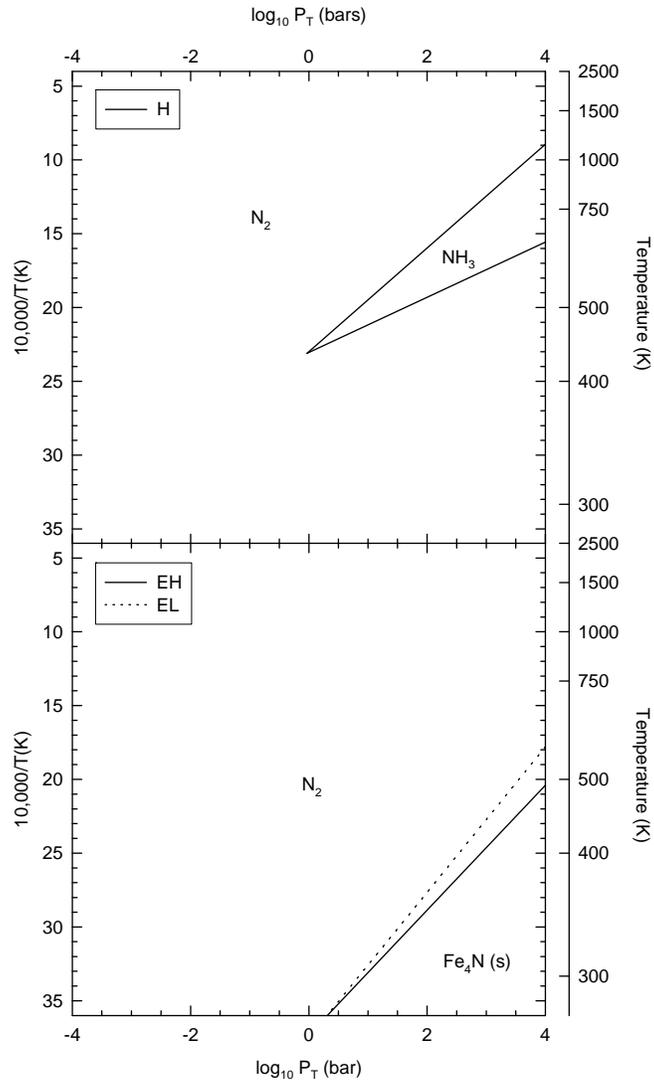

Figure 6: Nitrogen ratios in (a) H, (b) EH, (c) EL chondrites

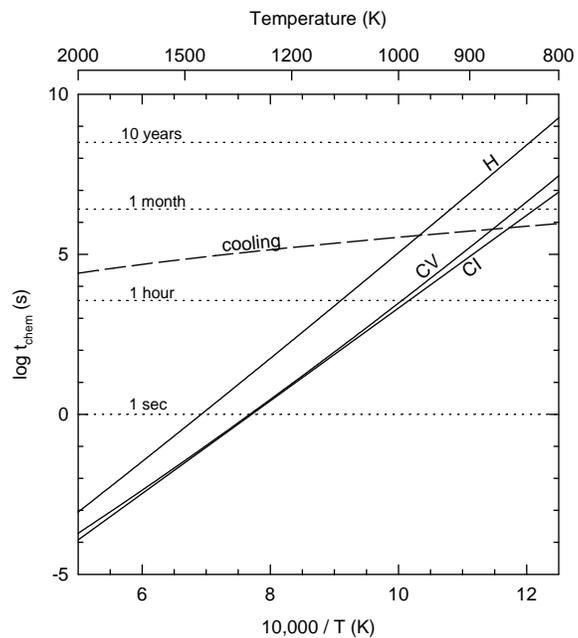

Fig. 7. Chemical lifetime of CO from the reaction $CO + OH = CO_2 + H$.